\documentclass{PoS}

\usepackage{amsmath}
\usepackage{amssymb}
\usepackage{epsfig}
\usepackage{graphicx}
\usepackage{psfrag}

\title{Dark matter variations ...}

\ShortTitle{Dark matter variations...}

\author{\speaker{Jean-Marie Fr\`{e}re}\thanks{Work funded by IISN and Belspo IAP}\\
        Service de Physique Th\'{e}orique, Universit\'{e} Libre de Bruxelles, Bruxelles\\
        E-mail: \email{frere@ulb.ac.be}}

\abstract{In this short presentation, we remind of significant unknowns regarding the distribution of Dark Matter  in our immediate neighborhood, and review the recent improvements in the obtained  limits on its abundance.}

\FullConference{Proceedings of the Corfu Summer Institute 2014 "School and Workshops on Elementary Particle Physics and Gravity",\\
		3-21 September 2014\\
		Corfu, Greece }

\begin{document}

\section{Introduction: the standard lore}
While the need for Dark Matter is strongly established from the observation of galactic and halo rotation curves, as well as even more directly
from gravitational lensing, its nature on one hand, and its small-scale distribution on the other are still largely unknown.

What is best known is the relative mass contribution of Dark Matter (DM) in large structures, like galactic halos.

In our galaxy, a spherical halo leads to the generally accepted  average  density of $  \approx 0.3 GeV /cm^3$ at the Sun's location. (In the tradition of particle physics, we set $c=1$, $1 GeV * c^2 = 1.78  10^{-24} g$

As far as the nature of DM, most of the approach is guided by the search for the simplest possible description, in absence of more information.
Probably the most conservative approach (Jupiter-like objects) has been discarded after micro-lensing observations failed to observe
a significant number of candidates in our Galaxy's halo.
The next simplest approximation consists in only one new stable or at least long-lived elementary particle (or a new particle and a corresponding mediator), and
even in this case, many models exist; notably based on R-parity conservation in SuperSymmetry, which leads to a stable Lightest Susy Particle or more recently relying on various "portals", where the interaction between ordinary matter and DM is mediated for example by the scalar (Brout-Englert-Higgs) boson sector, or some extra gauge boson with weak mixing with the visible sector.

The only established part about DM is its gravitational interaction...it is the only way in which it has been "inferred". Common prejudice has
it that it should be weakly interacting elementary particles. This is merely based on the fact that, for a cooling Universe a single DM particle with weak interaction provides for the end of DM annihilation and its decoupling in a way leading to  the currently observed density. It might of course be a mere coincidence, but typically provides a mass/interaction correlation, particularly if only one type of particle is involved.

Large numerical simulations have been used to model the DM evolution. While they reproduce the general features of our Universe, they
often present significant (fractal-like) structure at all scales. This takes raises the question of the "clumpiness" of Dark Matter even inside
our own galaxy.

\section{Dark matter disk? }
Even in the context of a "conventional" halo, the possibility exists that interaction with the visible disk leads to the formation of
a DM disk, rotating at a comparable speed. Even a very modest contribution to such a disk can affect considerably the prospects
for indirect DM detection through annihilation in the Sun or Earth and subsequent observation of the products, notably in large neutrino detectors \cite{Ling:2009cn},\cite{Ling:2009eh}.

\section{Dark matter in the Solar system? }

How much Dark Matter is there in the Solar system? Can it be detected, notably on gravitational grounds?

We need to distinguish here between the average halo density (which contributes to the motion of the Sun and its planetary system inside
the galaxy), with the usual $0.3 GeV/cm^3$ canonical value, and a possible local over-density, which we will now describe.
We will not discuss here the possible excess of Dark Matter around the planets themselves \cite{Adler:2009ir}.

We take here a very pragmatic (I would say purely empirical) attitude: we know little about dark matter, and if we can use observations
to constrain its distribution in any way, we should seize the opportunity.

It may of course be argued that such local over-density is improbable, for instance that the probability of capture of DM by the Solar System
is weak (such an argument depends of course on the actual distribution of the DM in the halo, and possibly on the presence of a DM disk),
this does not seem sufficient reason to refrain from using constraints from already available observations. Furthermore, the "clumpy" nature
of DM distributions could even plead for a pre-existing local excess, in which the Solar System would have formed.

Testing for the presence of a Dark Matter excess around the Sun is in principle straightforward, but must rely intrinsically on measurements
at (at least) two different distances from our star. It also requires the elimination of all other perturbations to the planets orbits, like non-sphericity of the Sun, perturbations from other planets or the asteroid belt. For lack of a global fit including those perturbations and the possible DM excess, the latter is for the time being deduced from the error bars of the standard fit. This is justified in that the DM effect would be a continuous, secular one as opposite to the more time-dependent effects.

Consider indeed a continuous DM distribution, depending only on the distance from the Sun. The attractive mass seen by 2 planets at different distance will depend upon the Gravitational constant, the reduced mass, calculated using the planet's mass $m$, the Solar mass $M_\odot$, and the Dark Matter  mass inside  the (nearly circular) orbit.
The tricky point here is that, while the planet's individual masses are determined independently from their own (natural or artificial) satellites \cite{Standish:1995}, the Solar mass is only fixed via the planetary motions. For an individual planet in circular orbit, the DM effect could thus, up to a point, be
absorbed in the definition of the Solar mass through  $K_0 \equiv G(M_\odot + m)$. 
This makes it obvious that information on the DM distribution can only be gathered by considering motion at different distances from the Sun.
The most obvious way is to compare the orbits of 2 planets. This indeed lead to the best limits for DM distribution between Earth and Mars (see below).
Another variable which is sensitive to the distribution of Dark Matter is the precession of the perihelion: it can only be defined in the case of elliptical orbits (even it the limit of small ellipticity is taken), and thus probes the radial dependance of the gravitational potential.

In ref. \cite{Frere:2007pi} , we took the point to study the modifications of  $K, E$ and $ L$ needed to maintain the orbital parameters, namely the orbit radius $a$, the excentricity $e$ and the period, $T$, in the presence of Dark Matter and the resulting modifications to the gravitational potential..
We also commented on previous work, notably the precise data from \cite{Pitjeva:2005} and the approaches of  \cite{Khriplovich:2006sq},\cite{Iorio:2006cn},\cite{Sereno:2006mw}.

The canonical values
\begin{eqnarray}
K_0 &=& 4 \pi^2 \frac{a^3}{T^2}~,
\label{kepler3}\\
E_0 &=& \frac{K_0}{2a}~,\\
L_0^2 &=& K_0 p~,
\end{eqnarray}

are then modified to

\begin{eqnarray}
E&=&E_0 + \Delta E~,\\
L^2&=& L_0^2 + \Delta L^2~,\\
K&=& K_0 + \Delta K~,
\end{eqnarray}

We took the example of a power-law distribution
\begin{equation}
\rho (r) = \rho_0 \left(\frac{r}{r_0}\right)^{-\gamma}~,
\end{equation}
where $r_0$ corresponds to the Earth's orbit
and obtained the corresponding variations of the $K$ constant and the orbital precession
(notations are fully detailed in ref \cite{Frere:2007pi}).

\begin{eqnarray}
\Delta K (\rho_0,\gamma) &=& -(4-\gamma) G M(a)~, \label{gammashift1}\\
\Delta \Theta(\rho_0,\gamma) &=& -\pi(3-\gamma) \frac{M(a)}{M_{\odot}}~.
\label{gammashift2}
\end{eqnarray}

\begin{figure}
\begin{center}
\psfrag{g}[c][c]{\tiny $\gamma$}
\psfrag{rho}[c][c]{\tiny $\log_{10}\rho_0^{max}~({\rm GeV/cm^3})$}
\psfrag{peri}[c][c]{\tiny $\Delta \Theta$ constraint}
\psfrag{K}[c][c]{\tiny $\Delta K$ constraint}
\psfrag{i}[c][c]{\tiny  new Saturn constraint}
\includegraphics[height=.45\textwidth]{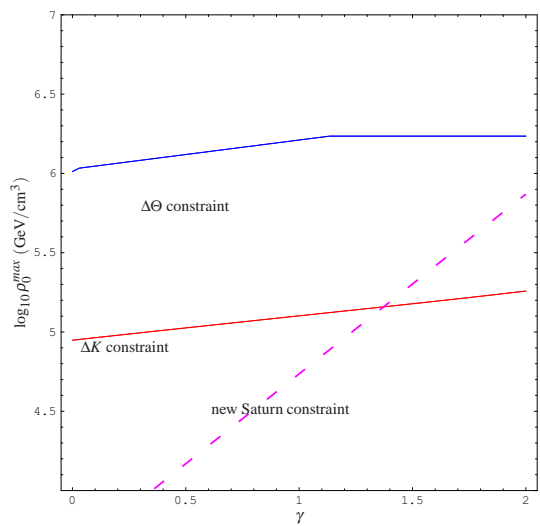}
\caption{Maximal allowed Dark Matter density at Earth's location, as a function of the halo profile parameter $\gamma$ from ref.\cite{Frere:2007pi} , the dashed curved
corresponds to the new limits using  \cite{Iorio:2013ida} converted to density at Earth}
\label{fig1}
\end{center}
\end{figure}

The comparison between those constraints is given in Fig. \ref{fig1}.
and  led us to a limit on the extra DM density at Earth of the order of $10^5 GeV/cm^3$. \\

Since this work, a general agreement has appeared on the calculation of the perihelion shift \cite{Iorio:2013ida} ,  but the most decisive 
progress has been in the determination of Saturn's orbital parameters due to the spatial mission Cassini , exploited by \cite{Pitjev:2013sfa} .

The improvement is significant, as \cite{Iorio:2013ida} quotes densities (at Saturn's distance) of  $ (5< \rho (at Saturn) < 8) \ 10^3 GeV/cm^3$ for a power-law exponent
$0 \leq \gamma \leq 4$, close to the values obtained by \cite{Pitjev:2013sfa} .

For the sake of comparison with the previous constraints, we should re-express these values in term of the densities at Earth used in \cite{Frere:2007pi}

The comparison is made in the Fig. \ref{fig1}, where we have included the previous upper bounds based on $K$, the slightly less stringent ones based on the perihelion shifts, and the latest values from \cite{Iorio:2013ida}

It should be noted that these limits completely exclude a significant contribution to the Pioneer anomaly, which would require a DM matter 
contribution of $8 \ 10^9 GeV/cm^3,  \gamma=0$ to account for the deceleration of the probes. 

It is also interesting to note as \cite{Pitjev:2013sfa}, that the total amount of DM allowed for withing Saturn's orbit is of the order of the current uncertainty on the Astero\"{\i}d belt mass  ($ \pm 2 \ 10^{-10} M_{Sun}$), which is not too surprising since Saturn brings a significant part of the constraint. Despite this smallness, the overdensity at Earth's distance could thus exceed the assumed Galactic density at the Sun by more than 4 orders of magnitude. 

Such measurements are important in terms of DM searches, which would be influenced by the lower relative velocity and easier capture of
a local excess by the Sun and planets. 

\section{Acknowledgements}
It is a pleasure to thank the organizers of the Corfu Summer Institutes for the beautiful meeting. This work received financial support from IISN (Belgium) and the Belgian Science Policy office (IAP program).

\end{document}